# Intrinsic defect properties in halide double perovskites for optoelectronic applications

Tianshu Li,[1] Xingang Zhao,[1] Dongwen Yang,[1] Mao-Hua Du,[2,*] and Lijun Zhang[1,*]

[1]State Key Laboratory of Superhard Materials, Key Laboratory of Automobile Materials of MOE, and College of Materials Science and Engineering, Jilin University, Changchun 130012, China.

[2]Materials Science and Technology Division, Oak Ridge National Laboratory, Oak Ridge, Tennessee 37831, United States.

(Dated: August 12, 2018)

Lead-free halide double perovskites with the formula of quaternary $A_2^+B^+B'^{3+}X_6^-$ have recently attracted intense interest as alternatives to lead-halide-perovskite-based optoelectronic materials for their non-toxicity and enhanced chemical and thermodynamic stability. However, the understanding of intrinsic defect properties and their effects on carrier transport and Fermi level tuning is still limited. In this paper, we show that, by exploring the phase diagram of a halide double perovskite, one can control the effects of intrinsic defects on carrier trapping and Fermi level pinning. We reveal the ideal growth conditions to grow p-type $Cs_2AgInCl_6$ and $Cs_2AgBiCl_6$ as well as semi-insulating $Cs_2AgBiBr_6$ with low trap density for targeted photovoltaic or visible-light/radiation detection application.

Lead (Pb) halide perovskites have been extensively investigated for diverse applications, including photovoltaics [1-10], light-emitting diode [11-13], laser [14-16], and radiation detection [17-20], owing to their unique electronic and optical properties. However, the toxicity of Pb and the intrinsic material instability have hindered their development. Recently, inorganic Pb-free halide double perovskites (HDPs), which are a large class of quaternary compounds with a general formula of $A_2^+B^+B'^{3+}X_6^-$, have attracted great attention as alternatives to Pb halide perovskites [21-31]. In particular, $Cs_2AgInCl_6$ and $Cs_2AgBiX_6$ (X=Cl, Br), which have been successfully synthesized in experiment and have the bandgap values of 2.0-3.0 eV and good material stability, have shown great potential as useful optoelectronic materials such as photovoltaic (PV) absorbers [32-35], photon and ionizing radiation detectors [36,37]. $Cs_2AgInCl_6$ exhibits direct band gap (2.0 eV according to the photoluminescence measure) as well as an ultra-long carrier lifetime (6 μs) [21,38-40], which are suitable for PV applications. However, Meng *et al.* reported that the optical absorption at the visible range in $Cs_2AgInCl_6$ may be significantly reduced due to the parity-forbidden transition at band edges [41]. Nevertheless, the low trap density (towards $10^8$ cm$^{-3}$) in single-crystal $Cs_2AgInCl_6$ makes it a promising ultraviolet detector material [37], which may find applications in fire and missile flame detection as well as optical communications. $Cs_2AgInBr_6$, with a theoretically predicted direct band gap of ~1.50 eV [21], is expected to be a good PV absorber or visible-light detector, although its synthesis is still a challenge. In contrast to the direct band gaps found in $Cs_2AgInX_6$, $Cs_2AgBiCl_6$ and $Cs_2AgBiBr_6$ have indirect band gaps of 2.77 and 2.19 eV, respectively [25,29,42]. Despite their relatively large band gaps for PV applications, it has been reported that the solar cells based on $Cs_2AgBiBr_6$ films show power conversion efficiencies up to 2.5% without device optimization [43]. The efficiency might be further improved by narrowing the bandgap of $Cs_2AgBiBr_6$ via trivalent metal doping/alloying [31,44]. Moreover, the indirect band gap is desirable for a semiconductor photon/radiation detection material because a direct band gap would increase the rate of radiative recombination of the radiation-generated electrons and holes, which should be collected by electrodes. The heavy constituent atoms and the sufficiently large band gap renders $Cs_2AgBiBr_6$ a potential X-/gamma-ray detector material. Indeed, a recent experimental work demonstrated promising figure of merits of $Cs_2AgBiBr_6$ for X-ray detection [36].

To advance the development of Pb-free HDPs as PV or photon/radiation detection materials, comprehensive understanding of intrinsic defect properties is of vital importance because defects strongly affect carrier density and transport. Efficient carrier transport is critically important in both PV and photon/radiation detection applications. High resistivity is required for semiconductor photon/radiation detection materials whereas good n- or p-type conductivity is usually needed in PV materials. Therefore, the understanding of carrier trapping at defects and carrier compensation is important for the development of HDPs for these optoelectronic applications. In this work, using advanced first-principle calculations, we identify deep defect levels that are detrimental to carrier transport in $Cs_2AgInCl_6$, $Cs_2AgBiCl_6$, and $Cs_2AgBiBr_6$ and show that chemical potentials of the constituent elements (which are associated with experimental growth conditions) can be modified to suppress unwanted deep level defects and to tune Fermi level for PV or photon/radiation detection applications.

We performed first-principles calculations based on density-functional theory, as implemented in the Vienna ab initio simulation package codes [45]. The electron-core interaction was described using the frozen-core projected augmented wave pseudopotentials [45,46]. The kinetic energy cutoff of 400 eV was used in all calculations. The lattice parameters and atomic positions were optimized using the Perdew-Burke-Ernzerhof (PBE) exchange-correlation functional [47]. The band gaps and the defect formation energies were further calculated using the screened Heyd-Scuseria-Ernzerh (HSE) hybrid density functional [48,49] with the spin-orbital coupling (SOC) based on the PBE-optimized structures without further relaxation. The primitive cell and a $4\times4\times4$ k-point mesh were used for obtaining the lattice parameters and band gaps while an 80-atom supercell was used for defect calculations. A $2\times2\times2$ and a $\Gamma$-point-only k-point mesh were used for the PBE and HSE supercell calculations, respectively. The mixing parameters of 25%, 28% and 30% for $Cs_2AgInCl_6$, $Cs_2AgBiCl_6$ and $Cs_2AgBiBr_6$ were used in HSE calculations for more accurate description of electronic structures. The calculated lattice parameters and band gaps of $Cs_2AgInCl_6$, $Cs_2AgBiCl_6$ and $Cs_2AgBiBr_6$ agree well with the experimental results (see TABLE S1, Supplemental Material [50]). The calculations of defect formation energies and transition levels were performed with the standard supercell approach, as described in detail in Supplemental Material [50].

The crystal structure of a HDP is shown in FIG. 1a. The $BX_6$ and $B'X_6$ octahedra alternate along the three crystallographic axes such that the B and B' ions form a rock-salt-type ordering. The band structures and the density of states of $Cs_2AgInCl_6$, $Cs_2AgBiCl_6$, and $Cs_2AgBiBr_6$ are shown in FIGs. 1(b)-(c) and 5(a). $Cs_2AgInCl_6$ exhibits a direct band gap at the $\Gamma$ point; the conduction band is derived from the In-5$s$ states while the valence band states are mainly made up of Ag-4$d$ and the Cl-3$p$ states. In contrast, $Cs_2AgBiCl_6$, and $Cs_2AgBiBr_6$ have indirect band gaps; the conduction band minimum (CBM) is located at the $\Gamma$ point while the VBM is at the X point. The Bi-6$p$ states dominate the conduction band. For the valence band, in addition to the contribution from the Ag-4$d$ and the halogen-$p$ states, the fully-occupied Bi 6$s$ states also hybridize with the valence band states. The coupling between the Bi-6$s$ states and the directional Ag-4$d$ states along the (100) direction moves the VBM from the $\Gamma$ point to the X point [22]. Note that the SOC strongly modifies the electronic structure of the Bi compounds [FIGs. 1(c) and 5(a)] but has negligible effect on that of the In compound [FIG. 1(b)].

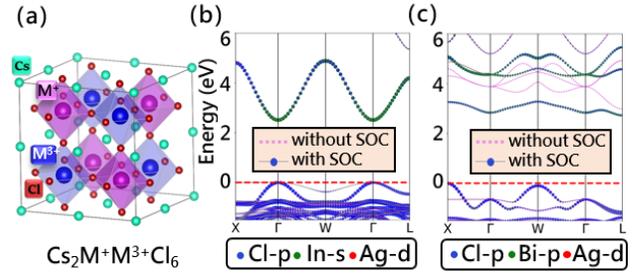

FIG. 1. (a) Double-perovskite crystal structure. Electronic band structure and orbital-projected density of states (b) $Cs_2AgInCl_6$ and (c) $Cs_2AgBiCl_6$.

As quaternary compounds, $Cs_2AgInCl_6$, $Cs_2AgBiCl_6$ and $Cs_2AgBiBr_6$ have many competing binary and ternary phases. We performed thorough evaluation of thermodynamic stability via the phase stability diagram analysis. A quaternary compound has three independent elemental chemical potentials; thus, the stable region for the single-phase quaternary compound is a polyhedron in a three-dimensional chemical potential space. Several cross-sections of the polyhedron corresponding to different values of the Ag chemical potential are shown for $Cs_2AgInCl_6$, $Cs_2AgBiCl_6$ and $Cs_2AgBiBr_6$ [FIGs. 2 and 5(b)]. Our results indicate that all three compounds can be grown in single phases. The defect formation energies for the three HDPs were calculated using the elemental chemical potentials corresponding to the selected points in the phase diagram to evaluate the tunability of the defect concentrations and the Fermi level.

There are a large number of intrinsic point defects in quaternary HDPs. Vacancies are generally important in halides due to their low formation energies. On the other hand, the close-packed perovskite crystal structure usually causes high formation energies for metal interstitials unless the metal ion has a low charge state of +1 [56]. Antisite defects in a multinary compound could be important except those due to substitution by an isovalent ion (i.e., $A^+$ on $B^+$ or $B^+$ on $A^+$), which leads to an electrically inactive neutral defect, or by an ion with a drastically different oxidation state (i.e., $B'^{3+}$ on $X^-$ and $X^-$ on $B'^{3+}$), which usually incurs high energy cost. For double perovskite halides, the antisite defects due to the disorder on B and B' sites are likely abundant because of the same octahedral environment for both sites. The above considerations led us to investigate 12 intrinsic defects in each of $Cs_2AgInCl_6$, $Cs_2AgBiCl_6$, and $Cs_2AgBiCl_6$, including four vacancies ($V_{Cs}$, $V_{Ag}$, $V_{In}/V_{Bi}$ and $V_{Cl}/V_{Br}$), six antisite substitutions ($Cs_{Cl}/Cs_{Cl}$, $Cl_{Cs}/Br_{Cs}$, $Cs_{Bi}/Cs_{Bi}$, $In_{Cs}/Bi_{Cs}$, $Ag_{In}/Ag_{Bi}$, and $In_{Ag}/Bi_{Ag}$) and two interstitials ($Cs_i$ and $Ag_i$).

Since the formation energy of a point defect depends on the chemical potentials of the constituent elements, we evaluated defect formation energies using the chemical potentials corresponding to a large number of points that sample the phase diagram. Representative points in the phase diagram [FIGs. 2 and 5(b)] were chosen to demonstrate the tunability of the defect concentration and the Fermi level [FIGs. 3-5]. Our calculations show that $V_{Ag}$ is the most important acceptor defect while the halogen vacancy $V_{Cl}/V_{Br}$, the antisite defect $In_{Ag}/Bi_{Ag}$, and $Ag_i$ are important donor defects. The relative stability of the above three donor defects differ in the three compounds ($Cs_2AgInCl_6$, $Cs_2AgBiCl_6$, and $Cs_2AgBiBr_6$). On the other hand, $V_{Ag}$ is the most stable acceptor defect in all three compounds even under the Ag-rich conditions [e.g., $\Delta\mu_{Ag}$ = 0.0 eV in FIG. 3(a)].

The results in FIG. 4 and S5 show that $V_{Ag}$ is a shallow acceptor and $Ag_i$ is a shallow donor in $Cs_2AgInCl_6$, $Cs_2AgBiCl_6$ and $Cs_2AgBiBr_6$. The antisite defects $In_{Ag}$ (in $Cs_2AgInCl_6$) and $Bi_{Ag}$ (in $Cs_2AgBiX_6$) insert relatively deep $(+2/+)$ and the $(+/0)$ levels inside the band gap. The halogen vacancy is shallow in $Cs_2AgInCl_6$ [FIG. 4(a)] but deep in $Cs_2AgBiCl_6$ [FIG. 4(b)] and $Cs_2AgBiBr_6$ (FIG. S5). The calculated $(+/0)$ transition levels of the halogen vacancies in $Cs_2AgInCl_6$, $Cs_2AgBiCl_6$, and $Cs_2AgBiBr_6$ are 0.04 eV, 0.92 eV, and 0.35 eV below their respective CBM. The deeper $(+/0)$ level of $V_{Cl}$ in $Cs_2AgBiCl_6$ than in $Cs_2AgInCl_6$ may be understood by considering that $Cs_2AgBiCl_6$ has a larger band gap than $Cs_2AgInCl_6$; in addition, $Bi^{3+}$ is a larger ion than $In^{3+}$ and the Bi dangling bond is a Bi-$6p$ orbital pointing directly towards $V_{Cl}$, leading to strong hybridization between the Bi-$6p$ and the Ag-$5s$ orbitals at $V_{Cl}$ and consequently deeper electron trapping. Going from $Cs_2AgBiCl_6$ to $Cs_2AgBiBr_6$, the halogen vacancy size increases and consequently the Bi-Ag hybridization is weakened at the vacancy; therefore, the (+/0) level of $V_{Br}$ in $Cs_2AgBiBr_6$ (FIG. S5) is shallower than that in $Cs_2AgBiCl_6$ [FIG. 4(b)].

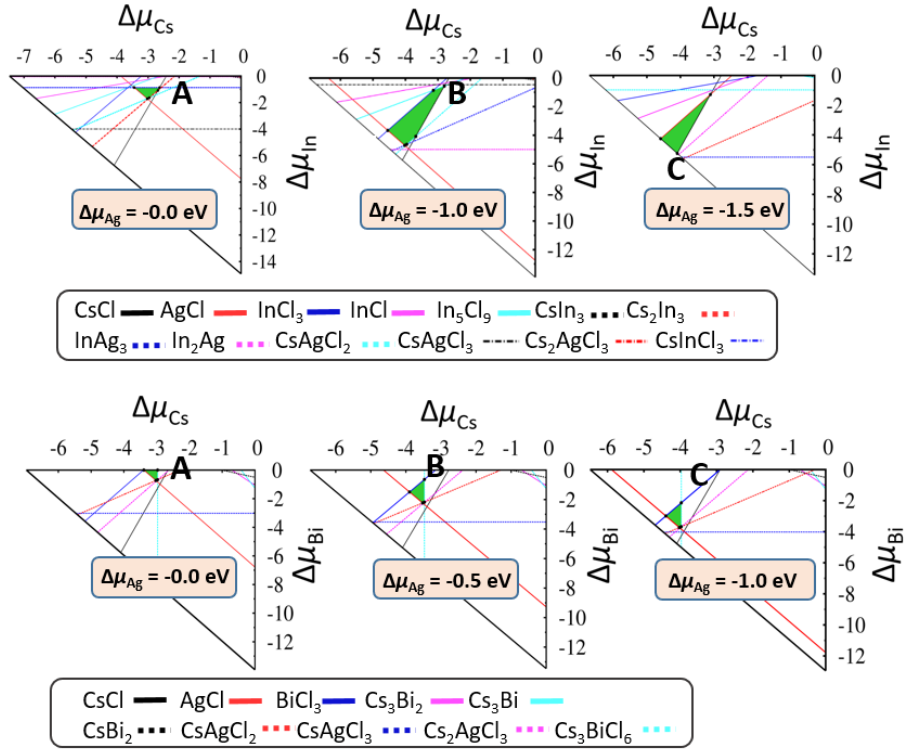

FIG. 2. Calculated phase diagrams of $Cs_2AgInCl_6$ (upper panel) and $Cs_2AgBiCl_6$ (lower panel) at different Ag chemical potential $\Delta\mu_{Ag}$. The green polygons indicate the stable regions of the single-phase HDP.

The electron trapping at halogen vacancies in the above three HDPs causes strong structural relaxation, which leads to the negative-U behavior in the case of $V_{Br}$ in $Cs_2AgBiBr_6$ with the (+/0) level above the (0/-) level. The $Ag^+$ and the $Bi^{3+}$ ions adjacent to the positively charged $V_{Br}^+$ repel each other. With successive addition of electrons to $V_{Br}^+$, the Ag-Bi distance is shortened to increase the hybridization and create a gap level that traps the electrons [57]. The calculated Ag-Bi distance at $V_{Br}$ in $Cs_2AgBiBr_6$ decreases from 5.90 Å, 3.03 Å, to 2.81 Å when the charge state of $V_{Br}$ is changed from +1, 0, to -1 (see FIG. S2). The Ag-Bi/In distance at $V_{Cl}$ in $Cs_2AgBiCl_6$ and $Cs_2AgInCl_6$ also shows the same trend (TABLE S2).

Among the low-energy defects in $Cs_2AgInCl_6$, only the antisite defect $In_{Ag}$ is a relatively deep donor defect [FIG. 3(a)]. This defect property is similar to that in $Cs_2AgInBr_6$ found in a previous DFT study [58]. In the two Bi HDPs, in addition to the antisite defect $Bi_{Ag}$, the halogen vacancy is also a deep donor [FIGs. 3(b) and 5(c)]. Compared to halide single perovskites, the halide double perovskites introduce additional antisite defects, which can induce deep gap states as shown above. The deep donor defects ($In_{Ag}$ in $Cs_2AgInCl_6$, $Bi_{Ag}$ and $V_{Cl}$ in $Cs_2AgBiCl_6$, and $Bi_{Ag}$ and $V_{Br}$ in $Cs_2AgBiBr_6$) are effective electron traps; thus, their concentrations need to be controlled. By sampling the chemical potentials in the phase diagram, we find that combining the In/Bi-poor and the halogen-rich conditions suppresses the concentrations of the above deep electron trapping defects, [see, for example, the point C in FIG. 2(a) for $Cs_2AgInCl_6$, the point C in FIG. S3 for $Cs_2AgBiCl_6$, and the point A in FIG. S4 for $Cs_2AgBiBr_6$]. Due to the low growth temperature of HDPs (below 210 ℃ for $Cs_2AgBiX_6$) [25],

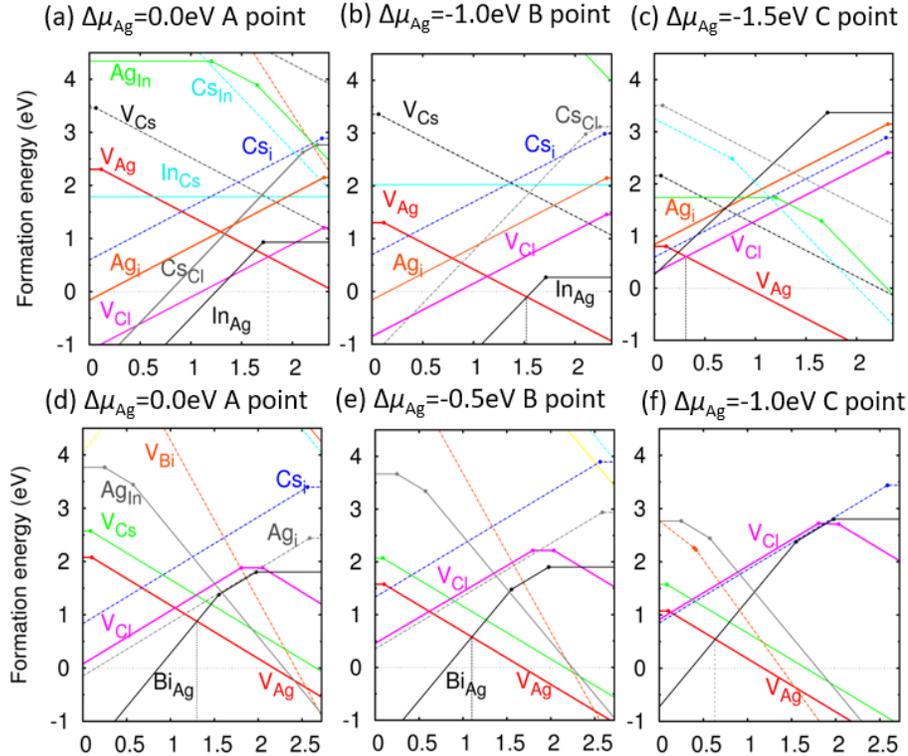

FIG. 3. The calculated formation energies of intrinsic defects in $Cs_2AgInCl_6$ (a, b, c) and $Cs_2AgBiCl_6$ (d, e, f), as a function of $E_F$, at three representative chemical potential points (A, B, and C) in FIG. 2. The slopes of the line segments indicate the defect charge

states and the kinks denote the transition energy levels. The Fermi levels are referenced to the host VBM. The defects with high formation energies are plotted using dash lines.

the formation energies of near 0.8 eV or higher [at the pinned Fermi level in FIG. 3(c) (point C), FIG. S3 (point C), FIG. S4(b) (point A)] should lead to low concentrations of the above deep donors according to Eq.S1. Thus, the In/Bi-poor and the halogen-rich growth conditions may lead to improved electron transport efficiency in $Cs_2AgInCl_6$ and $Cs_2AgBiX_6$.

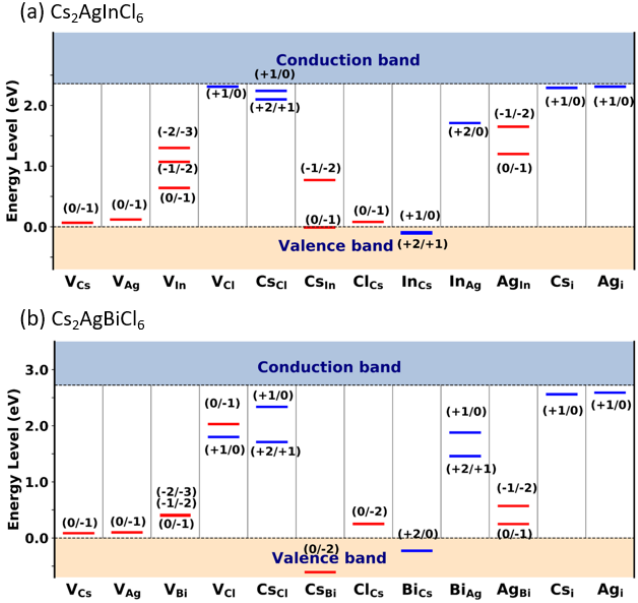

FIG. 4. The calculated transition energy levels for intrinsic donor (blue lines) and acceptor (red lines) defects in (a) $Cs_2AgInCl_6$ and (b) $Cs_2AgBiCl_6$, which are referenced to the host VBM (bottom dash lines).

The location of the Fermi level in a semiconductor is important for PV and radiation detection applications. Without external doping, the Fermi level can be pinned by the lowest-energy donor and acceptor defects. Our calculations show that the Fermi levels in the three HDPs can vary in a wide range within the band gap depending on the chemical potentials of the constituent elements [see FIGs. 3(a)-(c), S1 for $Cs_2AgInCl_6$, FIGs. 3(d)-(f), S3 for $Cs_2AgBiCl_6$, and FIGs. 5 and S4 for $Cs_2AgBiBr_6$]. In $Cs_2AgInCl_6$, the Cl chemical potential is most effective in tuning the Fermi level. $Cs_2AgInCl_6$ is $n$-type under Cl-poor conditions [FIGs. 3(a)-(b)] and $p$-type under Cl-rich conditions [FIG. 3(c)]. Although the Fermi level is too deep in $n$-type $Cs_2AgInCl_6$, the Cl-rich conditions can move the Fermi level reasonably close to the VBM [about 0.3 eV above the VBM as shown in FIGs. 1(c) and S1]. Thus, $p$-type $Cs_2AgInCl_6$ with low hole density may be attainable under the Cl-rich condition. Under this condition, the deep donor $In_{Ag}$ has relatively high formation energy as shown in FIG. 3(c). The other low-energy donor defect $V_{Cl}$ is shallow and thus is not an effective electron trap. Hence, $Cs_2AgInCl_6$ is potentially useful as a $p$-type solar absorber material with long minority carrier lifetime. Semi-insulating $Cs_2AgInCl_6$ may also be used as a radiation detector material.

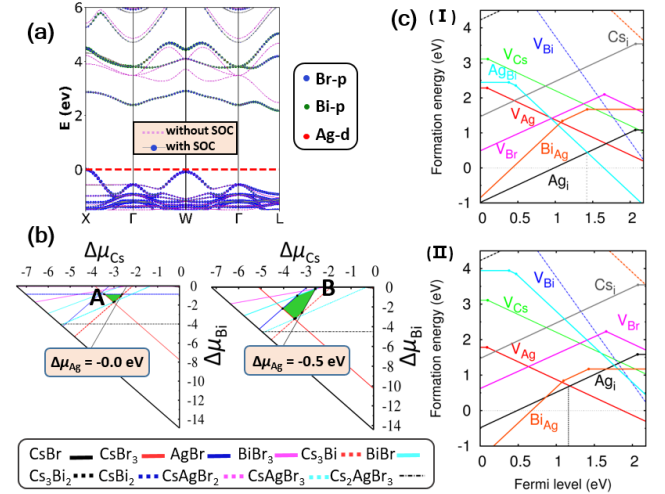

FIG. 5. (a) Electronic band structure and orbital-projected density of states of $Cs_2AgBiBr_6$. (b) Calculated phase diagram at $\Delta\mu_{Ag} = 0$ eV, and $\Delta\mu_{Ag} = -0.5$ eV. The green regions indicate the stable region for single-phase $Cs_2AgBiBr_6$. (c) The calculated formation energies of intrinsic defects in $Cs_2AgBiBr_6$, as a function of $E_F$, at the Point A (I) and the Point B (II) in the phase diagram (b). The slopes of the line segments indicate the defect charge states and the kinks denote the transition energy levels. The Fermi levels are referenced to the VBM of $Cs_2AgBiBr_6$.

The Fermi level of $Cs_2AgBiCl_6$ is confined within the $p$-type region based on our calculations [FIGs. 3(d)-(f) and S3]. The Cl-rich and the Bi-poor conditions (the point C in FIG. S3) can pin the Fermi level close to the VBM (0.18 eV above the VBM), leading to good $p$-type conductivity with high hole density. Under these conditions, the concentrations of the deep donors $Bi_{Ag}$ and $V_{Cl}$ are suppressed as discussed above. Therefore, similar to $Cs_2AgInCl_6$, $Cs_2AgBiCl_6$ may also be a potential $p$-type solar absorber material with long minority carrier lifetime. In contrast to $Cs_2AgInCl_6$ and $Cs_2AgBiCl_6$, the Fermi level of $Cs_2AgBiBr_6$ is more distanced away from both the CBM and the VBM (FIG. S4). This is consistent with the ob-served high resistivity in $Cs_2AgBiBr_6$ [36], which is required for a semiconductor radiation detection material. Since good carrier mobility and lifetime are also required for radiation detection, the combination of the Bi-poor and the Br-rich conditions (the point A in FIG. S4), which suppress the formation of the deep electron traps $V_{Br}$ and $Bi_{Ag}$, are desirable for $Cs_2AgBiBr_6$ detectors.

Using first principles calculations including spin-orbital coupling, we have systematically investigated the physical properties of defects in three representative HDPs, $Cs_2AgInCl_6$, $Cs_2AgBiCl_6$ and $Cs_2AgBiBr_6$ as promising optoelectronic materials. Detailed calculations of the formation energies and transition levels of intrinsic defects lead to the following important findings: (i) $Bi_{Ag}$ and halogen vacancies in $Cs_2AgBiCl_6$ and $Cs_2AgBiBr_6$ as well as $In_{Ag}$ in $Cs_2AgInCl_6$ are identified as deep electron traps with low formation energies and require Bi/In-poor and halogen-rich conditions to suppress; (ii) The Fermi level in three HDPs can be tuned by modifying the chemical potentials of their constituent elements (or equivalently the growth conditions). $Cs_2AgInCl_6$ and $Cs_2AgBiCl_6$ can both be p-type with the latter having higher attainable hole density while $Cs_2AgBiBr_6$ is usually semi-insulating with high resistivity; (iii) Based on the above results, $Cs_2AgInCl_6$ and $Cs_2AgBiCl_6$ may be potentially useful as *p*-type solar absorbers with good minority carrier lifetimes or as photon detectors while $Cs_2AgBiBr_6$ is a promising semiconductor radiation detection material. Our work provides valuable guidelines for further exploration of Pb-free perovskites for diverse applications.

## ACKNOWLEDGMENT

The authors acknowledge funding support from National Natural Science Foundation of China under Grant Nos. 61722403 and 11674121, the Recruitment Program of Global Youth Experts in China, and Program for JLU Science and Technology Innovative Research Team. Mao-Hua Du was supported by the U. S. Department of Energy, Office of Science, Basic Energy Sciences, Materials Sciences and Engineering Division. Calculations were performed in part at the high performance computing center of Jilin University.

* Mao-Hua Du: mhdu@ornl.gov

* Lijun Zhang: lijun_zhang@jlu.edu.cn

——————

# Supplemental Material for "*Intrinsic defect properties in halide double perovskites for optoelectronic applications*"

Tianshu Li,[1] Xingang Zhao,[1] Dongwen Yang,[1] Mao-Hua Du,[2,*] and Lijun Zhang[1,*]

[1]State Key Laboratory of Superhard Materials, Key Laboratory of Automobile Materials of MOE, and College of Materials Science and Engineering, Jilin University, Changchun 130012, China.

[2]Materials Science and Technology Division, Oak Ridge National Laboratory, Oak Ridge, Tennessee 37831, United States.

## I. Detailed description on defect calculations.

The probability of the formation of an intrinsic defect is determined by its formation energy. For a point defect $D$ in the charge state $q$, its equilibrium concentration in a crystal is given by

$$c(D,q) = N_{site} g_q \exp\left[-E^f(D^q)/k_B T\right] \quad (1)$$

where $k_B$ is the Boltzmann constant, $T$ is the temperature, $E^f(D^q)$ is the formation energy, $N_{site}$ is the number of possible atomic sites at which the defect may be formed, $g_q$ is the degeneracy factor for charge state $q$, which equals to the number of possible structural configurations and electron occupancies. The formation energy of a point defect can be evaluated as

$$E^f(D^q) = (E_D^q - E_h) - \sum_i n_i (\mu_i + \mu_i^{bulk}) + q(\varepsilon_{VBM} + \varepsilon_f) + \Delta E_{corr} \quad (2)$$

where $E_D^q$ is the total energy of the defect at the charge state $q$ in the supercell, $E_h$ is the energy of the defect-free supercell and $n_i$ is the difference in the number of atoms for the $i^{th}$ atomic species between the defect-containing and defect-free supercells. $\mu_i$ is the chemical potential of the $i^{th}$ atomic species relative to its bulk chemical potential $\mu_i^{bulk}$. $\varepsilon_{VBM}$ is the energy of valence-band maximum (VBM) of the defect-free material and $\varepsilon_f$ is the Fermi energy referenced to $\varepsilon_{VBM}$. The last term $\Delta E_{corr}$ corrects the error caused by the finite size of the supercell, including the image charge correction and the potential alignment correction [51-55]. The thermodynamic transition energy between two charge states $q$ and $q'$ can be determined by

$$\varepsilon(q/q') = \left[E^f(D,q';\varepsilon_f=0) - E^f(D,q;\varepsilon_f=0)\right]/(q-q') \quad (3)$$

when the Fermi level is below this energy, the charge state $q$ is stable; otherwise, the charge state $q'$ is stable.

For the HSE+SOC calculations of defects, we used only one k point (Γ point) due to the high computational cost. The convergence test performed at the PBE level shows that using a denser 2×2×2 $k$-point mesh increases the formation energies of the Cl vacancy ($V_{Cl}$) at the +1 and neutral charge states in $Cs_2AgBiCl_6$ by 0.11 eV and 0.15 eV, respectively, which results in an increase of the (+/0) transition level of $V_{Cl}$ by only 0.04 eV. This test shows that the error due to the $k$-point mesh is small and should not affect the conclusions of this paper.

## II. Supplementary Tables and Figures.

TABLE S1. Comparison of the crystal system, space group, lattice parameters and band gaps between calculated and experimental results for $Cs_2AgInCl_6$, $Cs_2AgBiCl_6$ and $Cs_2AgBiBr_6$.

|  | $Cs_2AgInCl_6$ | $Cs_2AgBiCl_6$ | $Cs_2AgBiBr_6$ |
|---|---|---|---|
| Crystal system |  | Cubic |  |
| Space group |  | $Fm\bar{3}m$ |  |
| a (Å) (Calc.) | 10.68 | 10.96 | 11.48 |
| a (Å) (Exp.) | 10.47[39] | 10.78[29] | 11.26[29] |
| Band gap (eV) (Calc.) | 2.36 | 2.72 | 2.18 |
| Band gap (eV) (Exp.) | 2.00[39] | 2.20-2.77[25,42] | 1.83-2.19[42] |

TABLE S2. The In/Bi-Ag distances at halogen vacancies of different charge states in $Cs_2AgInCl_6$, $Cs_2AgBiCl_6$ and $Cs_2AgBiBr_6$.

|  | $Cs_2AgInCl_6$ | $Cs_2AgBiCl_6$ | $Cs_2AgBiBr_6$ |
|---|---|---|---|
| Charge_+1 | 5.75 (Å) | 5.61 (Å) | 5.90 (Å) |
| Charge_0 | 5.65 (Å) | 3.48 (Å) | 3.03 (Å) |
| Charge_-1 | 2.74 (Å) | 2.82 (Å) | 2.81 (Å) |

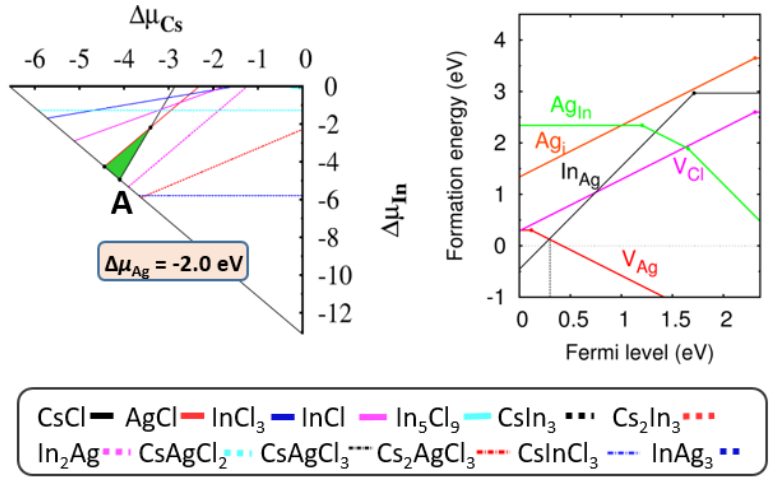

FIG. S1. (Left) Calculated phase diagram of $Cs_2AgInCl_6$ at $\Delta\mu_{Ag}$=-2.0eV. The green region indicates the stable region for the single-phase $Cs_2AgInCl_6$. (Right) The formation energies of dominant intrinsic defects as a function of $E_F$ at the Point A in the phase diagram in $Cs_2AgInCl_6$. The slopes of the line segments indicate the defect charge states and the kinks denote the transition energy levels. The Fermi levels are referenced to the host VBM.

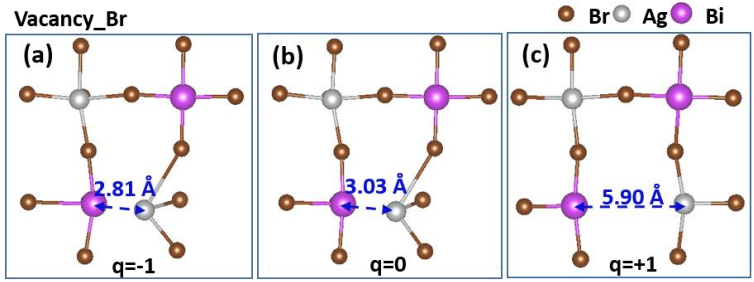

FIG. S2. The local structures of the Br vacancy at -1 (a), neutral (b), and +1 (c) charge states in $Cs_2AgBiBr_6$.

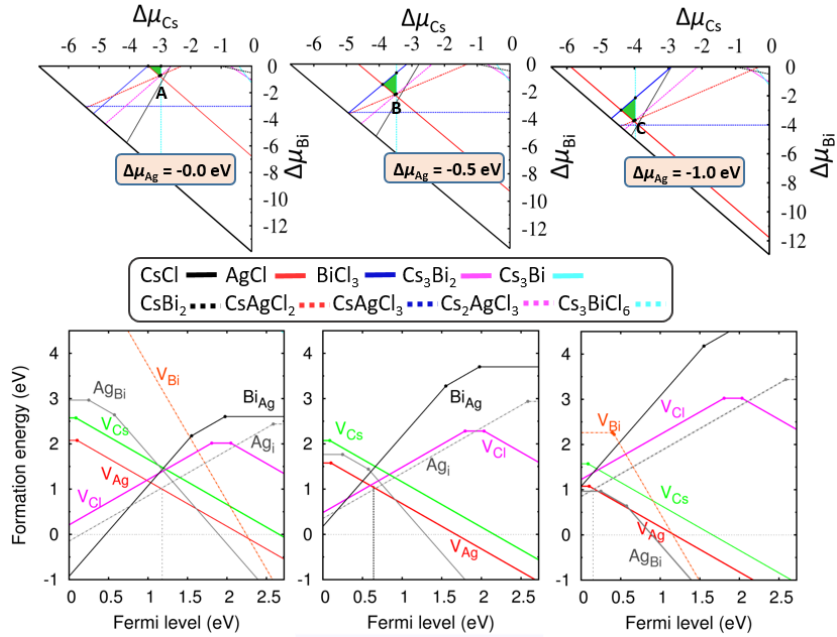

FIG. S3. The calculated formation energies of dominant intrinsic defects in $Cs_2AgBiCl_6$ at three chemical potential points A, B and C under Bi-poor growth conditions. The slopes of the line segments indicate the defect charge states and the kinks denote the transition energy levels. The Fermi levels are referenced to the host VBM

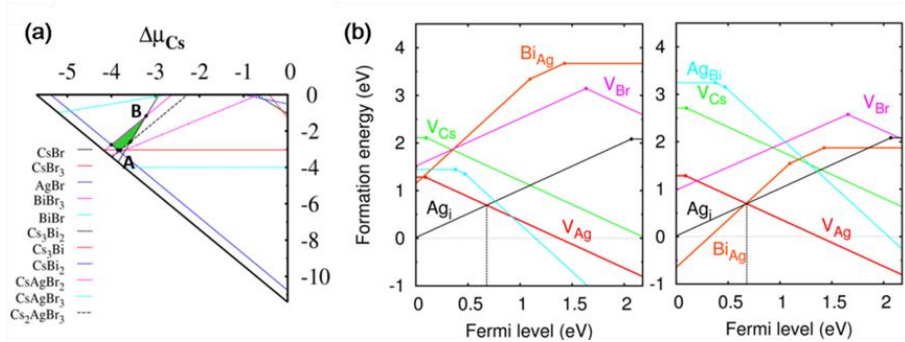

FIG. S4. (a) Calculated phase diagram of $Cs_2AgBiBr_6$ at $\Delta\mu_{Ag}$= -1.0eV. The green region indicates the stable region for the single-phase $Cs_2AgBiBr_6$. (b) The formation energies of dominant intrinsic defects as a function of $E_F$ at the Points A (left) and B (right) in

$Cs_2AgBiBr_6$. The slopes of the line segments indicate the defect charge states and the kinks denote the transition energy levels. The Fermi levels are referenced to the host VBM.

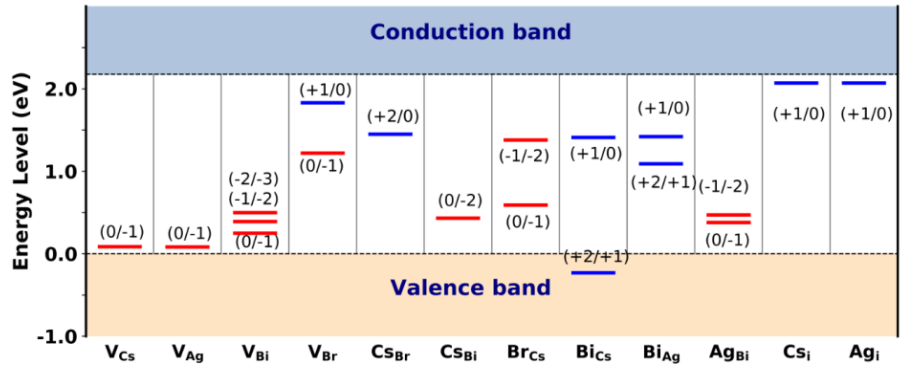

FIG. S5. The calculated transition energy levels for intrinsic donor (blue lines) and acceptor (red lines) defects in $Cs_2AgBiBr_6$, which are referenced to the VBM (bottom dash lines).

**References** (with the numbers consistent with those in the main text)